\documentclass[aps,preprint,amsmath,amssymb]{revtex4}
\usepackage{graphicx,color}

\def\g5{\gamma_{5}}
\def\ga{\gamma}

\def\e{\epsilon}

\def\be{\begin{eqnarray}}
\def\ed{\end{eqnarray}}
\def \t{\tilde}
\def\non{\nonumber}

\def\la{\langle}
\def\ra{\rangle}
\def\bs{\boldsymbol}
\def\lt{\lambda_t}

\begin{document}
\title{\Large \bf
CP phase of nonuniversal $Z'$ on $\sin\phi^{J/\Psi \phi}_s$ and\\
T-odd observables of $\bar B_{q}\to V_{q} \ell^{+} \ell^{-}$ }
\date{\today}

\author{\bf  Chuan-Hung Chen$^{1,2}$\footnote{Email:
physchen@mail.ncku.edu.tw}
 }

\affiliation{ $^{1}$Department of Physics, National Cheng-Kung
University, Tainan 701, Taiwan \\
$^{2}$National Center for Theoretical Sciences, Hsinchu 300, Taiwan
}

\begin{abstract}
The evidence of a large CP phase has been shown by CDF and D{\O}
collaborations in the time-dependent CP asymmetry (CPA) of $B_s\to
J/\Psi \phi$ decay, where the nonvanished CPA clearly implies the
existence of a non-Kobayashi-Maskawa phase in $b\to s$ transition.
We study the new phases originated by the extra family dependent
$U(1)'$ model and examine their impact on $\hat{T}$ or $CP$
violating observables in $\bar B_q\to V_q \ell^{+} \ell^{-}$ decays
with $V_q=K^*\ (\phi)$. Adopting the constraints from the current
data of $\Delta m_s$ and the branching ratios for $\bar B_s\to \mu^+
\mu^-$ and $\bar B_q\to V_q\ell^+ \ell^-$, we find that
$\phi^{J/\Psi\phi}_s=-(0.26\pm 0.22)$ and the T-odd observables of
above $10\%$ in the decay chain $\bar B_q\to V_q(\to K\pi [KK])
\ell^+ \ell^-$ can be achieved. In addition, we demonstrate that the
longitudinal and transverse polarizations of $K^*$ and the up-down
asymmetry of $K$ in the same decay chain are also sensitive to the
$Z'$-mediated effects.

\end{abstract}
\maketitle

Although three generations of the standard model (SM) provide a
unique CP violating phase in the Cabibbo-Kobayashi-Maskawa (CKM)
matrix \cite{CKM} to interpret the observed CP violating phenomena
in $K$ and $B$ systems, due to the failure of the KM phase in
the explanation of matter-antimatter asymmetry, some new source of CP
violation (CPV) is expected.
Therefore, it is an important
issue to probe and find the new CP violating effects. By the $B_s$
production in Tevatron Run II,
CDF and D{\O} have observed
the $B_s$ oscillation of $\Delta m_s=17.77 \pm 0.10 \pm 0.07$
ps$^{-1}$ \cite{CDF} and $\Delta m_s=18.56 \pm 0.87$ ps$^{-1}$
\cite{D0}, respectively, and the large direct CP asymmetry (CPA) of $0.39
\pm 0.17$ for $B_s\to K^{-} \pi^+$ \cite{BsCP}.
Moreover,
surprisingly
both detectors have also shown an unexpected large CP phase in the mixing-induced
CPA for $B_s\to J/\Psi \phi$
with the average value  given by
\cite{CDF-D0}
 \be
\phi^{J/\Psi \phi}_{s}=-2\beta_s + \phi^{\rm NP}_{s}= -2[0.10,
1.42]\label{eq:phis}
 \ed
at the $95\%$ confidence level (CL), where $\beta_s \equiv
arg(-V_{ts} V^*_{tb}/V_{cs}V^*_{cb})\approx 0.019$ \cite{CGL_PLB670}
is the CP phase of the SM while $\phi^{\rm NP}_s$ is a phase from  new physics.

If the measurement of Eq.~(\ref{eq:phis}) indicates the existence of
a new CP phase in $b\to s$ transition
\cite{CGL_PLB670,Botella:2008qm}, it is interesting to expose the
effects in other $b\to s$ related phenomena, such as the puzzles of
CPAs in $B\to K\pi$ and of polarizations in $B\to \phi K^*$.
Nevertheless, as known that owing to the inevitable large
uncertainty of nonperturbative QCD effects, it is difficult to
definitely uncover the new CP phase from the CPAs of the nonleptonic
exclusive decays. Therefore, the best environment to look for new
phase is that QCD effects are less involved and SM contributions are
highly suppressed. Therefore, we propose that the large CP violating
phase in $b\to s$ transition can be directly probed by measuring
T-odd observables in $\bar B_q \to V_q \ell^+ \ell^-$  decays with
$V_q=K^* (\phi)$.

To understand the property of $\hat{T}$ violating approach for
probing the new CP phase, we briefly discuss the characters of
CP-odd and T-odd observables. In a decay process, the direct CPA or
CP-odd observable is defined by ${\cal A}_{CP}\equiv(\bar\Gamma
-\Gamma)/( \bar \Gamma+\Gamma)$, where $\Gamma$ ($\bar{\Gamma}$) is
the partial decay rate of the (CP-conjugate) process. Due to ${\cal
A}_{CP}\propto\sin\theta_{W}\sin\theta_{S}$, it always needs weak CP
phase $\theta_{W}$ and strong phase $\theta_{S}$ simultaneously to
get the nonvanished CPA. Thus, the prediction efficiency on such CPA
is mainly dictated by the uncertain calculations of strong phase.
Another way to probe CP phase is through the spin-momentum triple
correlation of $\vec{s}\cdot ( \vec{p}_{i}\times \vec{p}_{j})$
\cite{geng-BRD,CG_NPB636}, where $\vec{s}$ is the spin carried by
one of outgoing particles and $\vec{p}_{i}$ and $\vec{p}_{j}$ denote
any two independent momentum vectors. The triple momentum
correlation is T-odd observable since it changes sign under the time
reversal ($\hat{ \rm T}$) transformation of $t\rightarrow -t$. We
note that $\hat{\rm T}$ is used to distinguish from real
time-reversal transformation which also contains the interchange of
initial and final states. By the CP$\hat{\rm  T}$ invariant theorem,
$\hat{\rm T}$ violation (TV) implies CPV. Therefore, study of the T-odd
observable can also help us to understand the origin of CPV.
Intriguingly,  the triple correlation
is proportional to
$\sin(\theta_{W}+\theta_{S})$. This leads that the strong phase in
T-odd observable is not necessary. To satisfy the criteria of
a small
QCD uncertainty with the high suppression in the SM for the $b\to s$
transition mentioned earlier, we find that TV in $B_q\to V_q\ell^{+}
\ell^{-}$ with polarized vector meson could provide a good chance to
directly observe the new phase
 revealed by CDF and D{\O}.

To search for the new CP phase in the extension of the SM, a model
independent analysis is not suitable for the current study.
In this
paper we investigate the minimal extension in the gauge sector by
including one family dependent $U(1)'$ to the SM gauge symmetry,
i.e. the considered gauge symmetry is $SU(3)_C\times SU(2)_L\times
U(1)_Y\times U(1)'_X$. Although lots of extensions of the SM predict
additional $U(1)'$ gauge symmetry, such as string theory \cite{String} ,
grand unified theories \cite{GUTs}, Little Higgs models \cite{LH},
light U-boson model \cite{Uboson}, etc, a well-motivated $Z'$
model for low energy systems is the so-called nonuniversal $Z'$.
Besides the $U(1)'$ charges carried by the fermions are family
dependent \cite{LP_PRD62}, the model also leads to flavor changing
neutral currents (FCNCs) at tree level and provides the solution to
the puzzles in $B$ physics \cite{BZ1}. The recent progress of
the nonuniversal $Z'$ on the phenomenological applications could be
referred to Refs.~\cite{BZ1,BZ2}.

We start to set up the relevant interactions with the new $Z'$ gauge
particle. Following the convention in
Refs.~\cite{LP_PRD62,LL_PRD45}, we write the couplings of the
$Z'$-boson to fermions as
 \be
{\cal L}^{Z'}&=& -g_2 J^{f\mu}_{Z'} Z'_{\mu}\,,
  \ed
with
 \be
J^{f\mu}_{Z'}&=& \bar F \ga^{\mu} \left[\bs{\t\e^L_f} P_L +
\bs{\t\e^R_f} P_R \right]F\,, \label{eq:jf}
 \ed
where $F^{T}=(f1, f2, f3)$ and  the indexes $fi$ (i=1,2,3) denote the
same type of quarks (qs) or charged leptons ($\ell s$),
$\bs{\t\e^\chi_f}$ are diagonal $3\times 3$ matrices in flavor space
with $\chi=L,\,R$ and are represented by $\rm
diag\bs{\t\e^\chi_f}=(X^{\chi}_{f1}, X^{\chi}_{f2}, X^{\chi}_{f3})$.
Note that $X^{\chi}_{fi}$ could be different for different
generations in the nonuniversal $Z'$ model.
Hereafter, $X^{\chi}_{fi}$ are taken as free parameters. Although
the third component of $SU(2)_L$ and $U(1)_Y$ in the SM also
provide two neutral gauge bosons,   we will not  show
them explicitly since they involve only
flavor conserved processes and are well known. To obtain the interactions in physical states, we
introduce $V^{\chi}_{f}$  to diagonalize the associated Yukawa
matrix of fermion. In terms of mass eigenstates, Eq.~(\ref{eq:jf})
could be written as
 \be
J^{f\mu}_{Z'}&=& \bar F \ga^{\mu} \left[ \bs{L^f} P_L + \bs{R^f} P_R
\right]F \,, \label{eq:intZp} \\
\bs L^f &=& V^{L}_{f} \bs{\t\e^L_f} V^{L\dagger}_f\,, \ \ \ \bs R^f
= V^{R}_{f} \bs{\t\e^R_f} V^{R\dagger}_f\,.\non
 \ed
Due to the unequal elements in $\bs{\t\e^{\chi}_f}$, obviously,
flavor misaligned effects in $\bs L^{f}$ and $\bs R^f$ will lead to
FCNCs at tree level.

In order to display the new physics effects with more compact form
in the processes, we use vector and axial-vector
currents instead chiral currents. Thus, Eq.~(\ref{eq:intZp}) is
rewritten by
 \be
J^{f\mu}_{Z'}&=& \bar F \ga^{\mu} \left[ \bs{c^{f}_{V} } +  \bs{
c^{f}_{A}} \ga_5
\right]F \,, \\
\bs{c^f_{V}} &=& \bs{R^{f}}+\bs{L^{f}}\,, \ \ \ \bs{c^f_{A}} =
\bs{R^{f}}-\bs{L^{f}}\,.\non
 \ed
Consequently, the  $Z'$-mediated interactions for $b\to s$ are given
by
 \be
{\cal L}^{Z'}_{b\to s} &=& -g_2 \bar s \ga^{\mu} \left( c^{q}_{Vsb}
+ c^q_{Asb} \gamma_5\right) b Z'_{\mu} + h.c. \label{eq:int_bs}
 \ed
For semileptonic decays, the couplings to the leptons are taken as
 \be
{\cal L}^{Z'}_{\ell\ell} &=& -g_2 \bar\ell \ga^{\mu}\left(
c^{\ell}_{V\ell\ell} + c^{\ell}_{A\ell\ell} \gamma_5\right) \ell
Z'_{\mu}\,. \label{eq:int_ll}
 \ed
Since we will concentrate on the $\ell^+\ell^-$-pair production,
to avoid the problem on the lepton flavor violation and
nonuniversality and  simplify our analysis, we assume that
$c^{\ell}_{V(A)\ell\ell}$ are the same for different families, i.e.
$c^{\ell}_{V(A)ee}= c^{\ell}_{V(A)\mu\mu}= c^{\ell}_{V(A)\tau\tau}$.
In the following analysis, we will show that the impacts of
$Z'$-mediated FCNCs on the processes of $B_s$ mixing, $\bar B_s\to
\mu^{+} \mu^{-}$ and $\bar B_q\to V_q \ell^{+} \ell^{-}$.  Note that
since we don't have any information on $V^{\chi}_{q}$ except the CKM
matrix defined by ${\bs V}=V^{L}_{U} V^{L\dagger}_{D}$, the elements
in $\bs c^{q}_{V, A}$ should be regarded as free parameters.
Therefore, the parameters associated with the processes for $b\to d$
and $s\to d$ transitions could be independent of those associated
with the $b\to s$ transition.

Based on the new effects, we now formulate the related
phenomena for $\Delta B=2$ and $\Delta B=1$ processes. In terms of
Eq.~(\ref{eq:int_bs}), the tree induced effective Hamiltonian for
$\Delta B=2$ is read by
 \be
{\cal H}^{Z'}_{\Delta B=2}= \frac{G_F \omega}{2\sqrt{2}} \left[\bar
s \ga_{\mu} \left( c^{q}_{Vsb} + c^{q}_{Asb} \ga_5 \right)b
\right]^2\,,
 \ed
where
$G_F/\sqrt{2}=g^2_1/8m^2_W$, $g_1$ is the gauge
coupling of $SU(2)_L$ and $\omega=4m^2_W(g_2/g_1)^2 /m^2_{Z'}$.
Including the Wick contracting factor and using the matrix elements
parametrized by
 \be
\la \bar B | \bar s \ga_{\mu} P_{\chi} b \bar s \ga^{\mu} P_\chi b |
B \ra &=& \frac{1}{3} m_B f^2_B \hat{B}_B\,, \non \\
\la \bar B | \bar s \ga_{\mu} P_{L} b \bar s \ga^{\mu} P_R b | B \ra
&=& -\frac{5}{12} m_B f^2_B \hat{B}_B\,,
 \ed
we obtain
 \be
M^{Z'}_{12} &=& \la \bar B | {\cal H}^{Z'}_{\Delta B=2} | B \ra\,,
\non
\\
&=&  \frac{G_F \omega}{24\sqrt{2}} m_B \hat{B}_{B} f^2_B
\left[-(c^q_{Vsb})^2 + 9 (c^q_{Asb})^2\right]\,.
 \ed
Combining the result with the SM contribution, the transition matrix
element for $\bar B_s \to B_s$ is given by
  \be
M^{s}_{12}&=& A^{\rm SM}_{12} e^{-2i\beta_s } + A^{Z'}_{12}
e^{2i(\theta^{Z'}_{s}-\beta_s)}\,, \label{eq:ms12}
 \ed
where $A^{Z'}_{12}=|M^{Z'}_{12}|$ and the CP phase of $M^{Z'}_{12}$
is parametrized by $2(\theta^{Z'}_{s}-\beta_s)$ \cite{CGL_PLB670}.
As a result, the $B_s$ mixing is obtained as
 \be
\Delta m_s = 2|M^{s}_{12}|=\Delta m^{\rm SM}_{s}\left(
1+2r\cos2\theta^{Z'}_{s} + |r|^2\right)^{1/2}
 \ed
with $r=A^{Z'}_{12}/A^{\rm SM}_{12}$. Due to $ \Delta \Gamma_s
\ll\Delta m_s$ in the B-system, the time-dependent CPA could be
simplified to be
 \be
-S_{J/\Psi\phi}&\simeq& {\rm
Im}\left(\sqrt{\frac{M^{s^*}_{12}}{M^{s}_{12}}}\right)
=\sin(2\beta_s - \phi^{Z'}_{s})\,, \non\\
\phi^{Z'}_{s}&=& \arctan\left(
\frac{r\sin2\theta^{Z'}_{s}}{1-r\cos2\theta^{Z'}_{s}} \right)\,.
 \ed

For $\Delta B=1$ processes, we focus on the important decay mode of
$b\to s \ell^{+} \ell^{-}$.
 From Eqs.~(\ref{eq:int_bs}) and
(\ref{eq:int_ll}), the corresponding effective interactions can be
found to be
 \be
{\cal H}^{Z'}_{b\to s\ell\ell} &=& \frac{G_F \omega}{2\sqrt{2}} \bar
s\ga_{\mu} \left(c^{q}_{Vsb} + c^{q}_{Asb}\ga_5 \right)b\;
 \bar\ell \ga^{\mu} \left(c^{\ell}_{V\ell\ell} +
c^{\ell}_{A\ell\ell}\ga_5 \right)\ell\,. \label{eq:bsll_int}
 \ed
The direct application of Eq.~(\ref{eq:bsll_int}) is the chiral
suppressed decay of $\bar B_s\to \ell^{+} \ell^{-}$. Using $\la 0 |
\bar s \ga_{\mu} \ga_5 b| \bar B_s (p)\ra=if_{B_s} p_{\mu}$ and
including the SM contributions, the branching ratio (BR) for $\bar
B_s\to \ell^{+} \ell^{-}$ is expressed by
 \be
{\cal B}(\bar B_s\to \ell^{+}
\ell^{-})=\tau_{B_s}\frac{G^2_F}{16\pi} m_{B_s}f^2_{B_s}m^2_{\ell}
\left( 1-\frac{4m^2_{\ell}}{m^2_{B_s}}\right)^{1/2}\left|A_{
\ell\ell} \right|^2\non
 \ed
with
 \be
A_{\ell\ell} &=& \frac{\lt\alpha}{\pi\sin^2\theta_W} Y(x_t)
+ \omega c^{q}_{Asb} c^{\ell}_{A\ell} \,, \non\\
\omega &=&\left(\frac{g_2}{g_1}\right)^2 \frac{4m^2_W}{m^2_{Z'}}\,,
 \ed
where $\lt=V^*_{ts} V_{tb}$, $\theta_W$ is the Weinberg angle,
$\alpha=e^2/4\pi$ and $Y(x_t)\approx 0.315 x^{0.78}_t$ with
$x_t=m^2_t/m^2_W$ \cite{BBL}.

To study the $\hat{T}$ violating effects in $\bar B_q\to V_q
\ell^+ \ell^-$ decays, besides the new interactions of
Eq.~(\ref{eq:bsll_int}),  we need the hadronic effects for the $\bar B_q
\to V_q$ transition. As usual,  the form factors associated with
various current operators are parametrized as
\begin{eqnarray}
\langle V(p_{V},\epsilon )| V_{\mu }| \bar{B}%
(p_{B})\rangle &=& i\frac{V(q^{2})}{m_{B}+m_{V}}\varepsilon _{\mu
\alpha \beta \rho }\epsilon ^{*\alpha }P^{\beta }q^{\rho },  \non \\
\langle V (p_{V},\epsilon )| A_{\mu }| \bar{B}
(p_{B})\rangle &=& 2m_{V} A_{0}(q^{2})\frac{\epsilon ^{*}\cdot q}{%
q^{2}}q_{\mu } \non \\
&& +( m_{B}+m_{V}) A_{1}(q^{2})\Big( \epsilon _{\mu
}^{*}-\frac{\epsilon ^{*}\cdot q}{q^{2}}q_{\mu }\Big)\non\\
&& -A_{2}(q^{2})\frac{\epsilon ^{*}\cdot q}{m_{B_q}+m_{V_q}}\Big( P_{\mu }-%
\frac{P\cdot q}{q^{2}}q_{\mu }\Big) ,  \non \\
\langle V(p_{V},\epsilon )| T_{\mu \nu }q^{\nu }| \bar{B}
(p_{B})\rangle &=& -iT_{1}(q^{2})\varepsilon _{\mu \alpha \beta \rho
}\epsilon ^{*\alpha }P^{\beta }q^{\rho },  \non\\
\langle V(p_{V},\epsilon )| T_{\mu \nu }^{5}q^{\nu }| \bar{B}
(p_{B})\rangle &=& T_{2}(q^{2})\Big( \epsilon _{\mu }^{*}P\cdot
q-\epsilon ^{*}\cdot qP_{\mu }\Big) \non \\
&& +T_{3}(q^{2})\epsilon ^{*}\cdot q\Big( q_{\mu
}-\frac{q^{2}}{P\cdot q}P_{\mu }\Big) \label{ffv}
\end{eqnarray}
with $V_{\mu }=\bar{s}\gamma _{\mu } b$, $A_{\mu }=\bar{s} \gamma
_{\mu }\gamma _{5} b$, $T_{\mu \nu }=\bar{s} i\sigma _{\mu \nu }b$,
$T_{\mu \nu }^{5}=\bar{s} i\sigma _{\mu \nu }\gamma _{5} b$,
$P=p_{B}+p_{V}$ and $q=p_{B}-p_{V}$.
Here, we have used $B$ ($V$) instead
of $B_q$ ($V_q$)
and $m_{B} (m_V)= m_{B_q}(m_{V_q})$.
Hereafter, we will adopt the notation for shorthand. Hence, the
transition amplitude for $\bar B\to V \ell^{+} \ell^{-}$ ($\ell=e,\,
\mu$) with the polarized $V$ can be arranged by \cite{CG_NPB636}
\begin{eqnarray}
{\cal M}_{V}^{(\lambda )} &=&-\frac{G_{F}\alpha \lambda_t}{2%
\sqrt{2}\pi }\left\{ {\cal M}_{1\mu }^{(\lambda )}L^{\mu }+{\cal
M}_{2\mu
}^{(\lambda )}L^{5\mu }\right\}\,,  \non \\
{\cal M}_{a\mu }^{(\lambda )} &=&i\xi_{1}\varepsilon _{\mu \nu
\alpha \beta }\epsilon ^{*\nu }(\lambda )P^{\alpha }q^{\beta
}+\xi_{2}\epsilon _{\mu }^{*}(\lambda )+\xi_{3}\epsilon ^{*}\cdot
qP_{\mu }\,,   \non
\end{eqnarray}
where  $L_{\mu}=\bar\ell \ga_{\mu} \ell$, $L^5_{\mu}=\bar\ell
\ga_{\mu}\ga_5 \ell$,  subscript $a=1[2]$ while $\xi_{i}=h_{i}$
$[g_{i}]$ ($i=1$, $2$, $3$) and their explicit expressions are
 \be
h_{1} &=&\left[C^{\rm eff}_9 +c^{q}_{Vsb} c^{\ell}_{V\ell\ell} \zeta
\right]\frac{V}{m_{B}+m_{V}}
+\frac{2m_{b}}{q^{2}}C_7T_{1}\,, \non \\
h_{2} &=&\left[-C^{\rm eff}_9 +  c^{q}_{Asb} c^{\ell}_{V\ell\ell}
\zeta\right](m_{B}+m_{V})A_{1} \non
P\cdot q C_7 T_{2}\, ,  \non \\
h_{3} &=&\left[C^{\rm eff}_9 -c^{q}_{Asb} c^{\ell}_{V\ell\ell}\zeta
\right] \frac{A_{2}}{m_{B}+m_{V}}
+\frac{2m_{b}}{q^{2}} C_7\Big(T_{2}+\frac{q^2}{P\cdot
q}T_{3}\Big)\,,
\non \\
g_{1}&=&\left[C_{10} + c^{q}_{Vsb} c^{\ell}_{A\ell\ell} \zeta\right]
\frac{V}{ m_{B}+m_{V}}, \non \\
g_{2} &=&\left[-C_{10}+c^{q}_{Asb} c^{\ell}_{A\ell\ell} \zeta \right]
(m_{B}+m_{V})A_{1}\,,\non \\
g_{3}
&=&\left[C_{10}-c^{q}_{Asb}c^{\ell}_{A\ell\ell}\right]\frac{A_{2}}{
 m_{B}+m_{V}}\,.
%
  \label{eq:ampk1}
 \ed
Here, we have suppressed the $q^2$-dependence in the form factors,
$(C^{\rm eff}_{9},\, C_7,\, C_{10})$ are the Wilson coefficients of
the SM and their specific results could be referred to
Ref.~\cite{BBL} and $\zeta=\pi \omega/\alpha \lt$.

To obtain the $\hat{T}$ violating obervables, which are connected to
the polarizations of the vector meson in $\bar B\to V \ell^+ \ell^-$
decays, we have to consider the decay chain $\bar B\to V(\to P_1
P_2)\ell^{+} \ell^{-}$ in which $P_1 P_2$ is $K\pi(KK)$ as
$V=K^*(\phi)$. For calculations, we set the positron lepton momentum
as $p_{\ell^+}=\sqrt{q^2}/2\,(1,
\sin\theta_{\ell^+},0,\cos\theta_{\ell^+})$ in the $q^2$ rest frame
and the K momentum $p_K =m_V/2 (1, \sin\theta_K \cos\phi_K,
\sin\theta_K \sin\phi_K, \cos\theta_K)$ in the $V$ rest frame. With
some tedious algebra, the differential decay rate with polarized V-meson
can be formulated by \cite{CG_NPB636,BHP}
\begin{widetext}
 \be
\frac{d\Gamma }{d\Omega_K dq^{2}} &=&\frac{
G_{F}^{2}\alpha^{2}|\lambda _{t}| ^{2}| \vec{p} |
}{2^{14}\pi ^{6}m_{B}^{2}}{\cal B}(K^*[\phi]\rightarrow K\pi[KK]) \nonumber \\
&&\times \Big\{16\cos ^{2}\theta _{K}\sum_{a=1,2}| {\cal M}_{a}^{0}|
^{2}  +8\sin ^{2}\theta _{K}\sum_{a=1,2}\Big( | {\cal
M}_{a}^{+}|^{2}+| {\cal M}_{a}^{-}|^{2}\Big) \nonumber \\
&&-8\sin ^{2}\theta _{K}\Big[ \cos 2\phi_K \sum_{a=1,2}{Re }{\cal
M}_{a}^{+}{\cal M}_{a}^{-*} +\sin 2\phi_K \sum_{a=1,2}{Im}
 {\cal M}_{a}^{+}{\cal M}_{a}^{-*}\Big]  \nonumber \\
&&+3\pi\sin 2\theta _{K}\Big[ \cos \phi_K \Big( {Re}{\cal M}
_{1}^{0}({\cal M}_{2}^{+*}-{\cal M}_{2}^{-*})+{Re}({\cal
M}_{1}^{+}-{\cal M}
_{1}^{-}){\cal M}_{2}^{0*}\Big)   \nonumber \\
&&+\sin \phi_K \Big( {Im}{\cal M}_{1}^{0}({\cal M}_{2}^{+*}+ {\cal
M}_{2}^{-*})-{Im}({\cal M}_{1}^{+}+{\cal M}_{1}^{-}){\cal
M}_{2}^{0*}\Big) \Big] \Big\}  \label{eq:difangle}
 \ed
 \end{widetext}
where $d\Omega_K=d\cos\theta_K d\phi_K$, $\theta_{K}$  is the polar
angle of K-meson while the $\phi_K$ is the relative angle of
decaying planes of $P_1P_2$ and $\ell^+ \ell^-$, $|\vec{p}
|=[((m_{B}^{2}+m_{V}^{2}-q^{2})/(2m_{B}))^{2}-m_{V}^{2}]^{1/2}$, $
{\cal M}_{a}^{0}$ and ${\cal M}_{a}^{\pm }$ denote the longitudinal
and transverse polarizations of $V$ respectively and they are given
as follows:
\begin{eqnarray*}
{\cal M}_{a}^{0} &=&\sqrt{q^{2}}\left( \frac{E_{V}}{m_{V}}\xi_{2}+2
\sqrt{q^{2}}\frac{\left| \vec{p}_{V}\right|
^{2}}{m_{V}}\xi_{3}\right)
, \\
{\cal M}_{a}^{\pm } &=&\sqrt{q^{2}}\left( \pm 2\left|
\vec{p}_{V}\right| \sqrt{q^{2}}\xi_{1}+\xi_{2}\right) \label{eq:Ma}
\end{eqnarray*}
with $|\vec{p}_{V}|=\sqrt{E^2_V-m^2_V}$ and
$E_V=(m^2_B-m^2_V-q^2)/2\sqrt{q^2}$. Since we have integrated out
the angles associated with the lepton, some important effect such as
the forward-backward asymmetry (FBA) of the lepton does not show up in
Eq.~(\ref{eq:difangle}). However, we find an up-down (UD) asymmetry
of the $K$-meson, which is  associated with $\sin2\theta_K$ and similar
to the lepton FBA, could be generated.

According to Eq.~(\ref{eq:difangle}), it is clear that besides the
real pieces, the imaginary parts that represent the TV (or CPV) also
appear in the angular distribution. If we can find the proper
$T$-odd operators denoted by ${\cal O}_{T_i}$, in terms of the
concept of quantum mechanics, the average of the $T$-odd operator
can be written as
 \be
\la {\cal O}_{T_i} \ra = \int d\Omega_K {\cal O}_{T_i} {\rm
sign}(\omega_i) \frac{d\Gamma}{d\Omega_K dq^2} \non
 \ed
where sign($\omega_i$) is chosen to get a nonvanished asymmetric
observables when integrating the angles of phase space.  Then, the
$T$-odd observable could be defined by
 \be
{\cal A}_{T_i}&=& \frac{\la {\cal O}_{T_i} \ra }{\int
d\Omega_K\left(d\Gamma/d\Omega_K dq^2 \right) }\,. \non
 \ed
Accordingly, we find that the two $\hat{T}$ violating terms in
Eq.~(\ref{eq:difangle}) can be extracted by using $T$-odd momentum
correlations, defined by
 \be
{\cal O}_{T_1}&=& 4 | \vec{p}_{B}| \frac{( \vec{p}_{B}\cdot
 \vec{p}_{\ell^{+}}\times \vec{p}_{K}) ( \vec{p}_{B}\times \vec{p}
 _{K}) \cdot ( \vec{p}_{\ell^{+}}\times \vec{p}_{B}) }{|
 \vec{p}_{B}\times \vec{p}_{K}| ^{2}| \vec{p}_{\ell^{+}}\times
 \vec{p} _{B}| ^{2}} \,,\non\\
 &=& 2\sin2\phi_K \,,\non \\
{\cal O}_{T_2}&=&\frac{16}{3\pi} \frac{|
\vec{p}_{B}|\vec{p}_{K}\cdot (
 \vec{p}_{B}\times \vec{p}_{\ell^{+}}) }{| \vec{p}_{B}\times
 \vec{p} _{K}| | \vec{p}_{B}\times \vec{p}_{\ell^{+}}| } =\frac{16}{3\pi}
 \sin\phi_K \non
 \ed
where the momenta of particles are taken in the $V$-meson reset
frame, ${\rm sign}(\omega_1)={\rm sign(\sin\theta_K)}$ and ${\rm
sign}(\omega_2)={\rm sign(\cos\theta_K)}$. As a result, we get
 \be
{\cal A}_{T_{1}}&=& -\frac{\sum_{a=1,2}Im({\cal M}_{a}^{+}{\cal
M}_{a}^{-*})}{  \sum_{\lambda=+,0,- }\sum_{a=1,2}|{\cal
M}_{a}^{\lambda }|^{2} }\,, \label{eq:AT1}
\\
{\cal A}_{T_{2}}&=& \frac{1}{\sum_{\lambda=+,0,- }\sum_{a=1,2}
|{\cal M}_{a}^{\lambda }|^{2} }\Big[{Im}{\cal M}_{1}^{0}({\cal
M}_{2}^{+*}+ {\cal
M}_{2}^{-*})
-{Im}({\cal M}_{1}^{+}+{\cal M}_{1}^{-}){\cal M}_{2}^{0*}
\Big]\,. \label{eq:AT2}
 \ed
Although our main purpose is to explore the existence of new CP
phase,
the $Z'$-mediated effects also have significant contributions to
CP-conserved quantities. Therefore, besides the physical observables
defined in Eqs.~(\ref{eq:AT1}) and (\ref{eq:AT2}), we could also
invent interesting measurable quantities, such as longitudinal and
transverse polarizations of the $V$-meson and the angular asymmetry
of the $K$-meson, etc \cite{CG_NPB636}. According to the definition
in Ref.~\cite{CG_NPB636},  the longitudinal and transverse
polarizations of $V$-meson appearing in Eq.~(\ref{eq:difangle}) can
be found as
 \be
 {\cal P}_{L}(q^2)&=& \frac{ \sum_{a=1,2}| {\cal M}_{a}^{0}| ^{2}}{
 \sum_{\lambda=+,0,- }\sum_{a=1,2}|{\cal M}_{a}^{\lambda
}|^{2}}\,, \label{eq:pl}  \\
{\cal P}_{T}(q^2)&=& \frac{ \sum_{a=1,2}| {\cal M}_{a}^{+}| ^{2}+|
{\cal M}_{a}^{-}| ^{2}}{  \sum_{\lambda=+,0,- }\sum_{a=1,2}|{\cal
M}_{a}^{\lambda }|^{2} }\,. \label{eq:pt}
 \ed
 In addition, due to the appearance of
$\sin 2\theta_K \cos\phi_K$ in Eq.~(\ref{eq:difangle}), the UD
asymmetry of the $K$-meson can be obtained as \cite{CG_NPB636}
 \be
{\cal A}^{K}_{UD}&=& \frac{ 1 }{\sum_{\lambda=+,0,-
}\sum_{a=1,2}|{\cal M}_{a}^{\lambda }|^{2} }\Big[{Re}{\cal M} _{1}^{0}
({\cal M}_{2}^{+*}-{\cal M}_{2}^{-*})
 +{Re}({\cal M}_{1}^{+}-{\cal M} _{1}^{-}){\cal
M}_{2}^{0*}\Big]\,.
 \label{eq:UDA}
 \ed

So far we have introduced  seven independent
free parameters of $\omega$, $|c^{q}_{Vsb}|e^{i\theta_V}$ and
$|c^{q}_{Asb}|e^{i\theta_A}$ and $c^{\ell}_{V\ell\ell}$ and
$c^{\ell}_{A\ell\ell}$
in
our analysis. In order to further reduce the number of
free parameters to simplify the phenomenological analysis, we
set $g_1=g_2$ and $m_{Z'}=1$TeV and force $\theta_V= \theta_A$.
Although the restricted condition of the latter is unnecessary,
it could correspond to a special scenario of Yukawa matrices,
e.g. hermitian Yukawa matrices which result in $V^L_{f}=V^R_{f}$
\cite{Hermitian}. For numerical estimations, we need the data to
constrain the remaining five parameters. Therefore, we adopt the
values of $\Delta m_s=18.17\pm 0.86$ ps$^{-1}$ \cite{CDF,D0},
${\cal B}(\bar B_d\to K^{[*]0}\mu^{+} \mu^{-})=
(0.45^{+0.12}_{-0.10})[1.05^{+0.15}_{-0.13}]\times
10^{-6}$~\cite{HFAG} and ${\cal B}(\bar B_s\to \mu^+ \mu^-)<4.7
\times 10^{-8}$~\cite{CDFBsmumu} as the inputs whereas the SM
predictions are taken as $\Delta m^{\rm SM}_s=17.37\pm 6.03$
ps$^{-1}$ \cite{LN}, ${\cal B}(\bar B_d\to K^{[*]0}\mu^{+}
\mu^{-})_{\rm SM}= 0.58[1.3]\times 10^{-6}$ \cite{CGA21} and ${\cal
B}(\bar B_s\to \mu^{+} \mu^{-})_{\rm SM}\approx 0.33\times 10^{-8}$
\cite{Bsmumu_th}. Since $\bar B_s\to \phi \ell^+ \ell^-$ is very
similar to $\bar B\to K^* \ell^+ \ell^-$,
we use the latter as
the delegate.

\begin{figure}[bpth]
\includegraphics*[width=4 in]{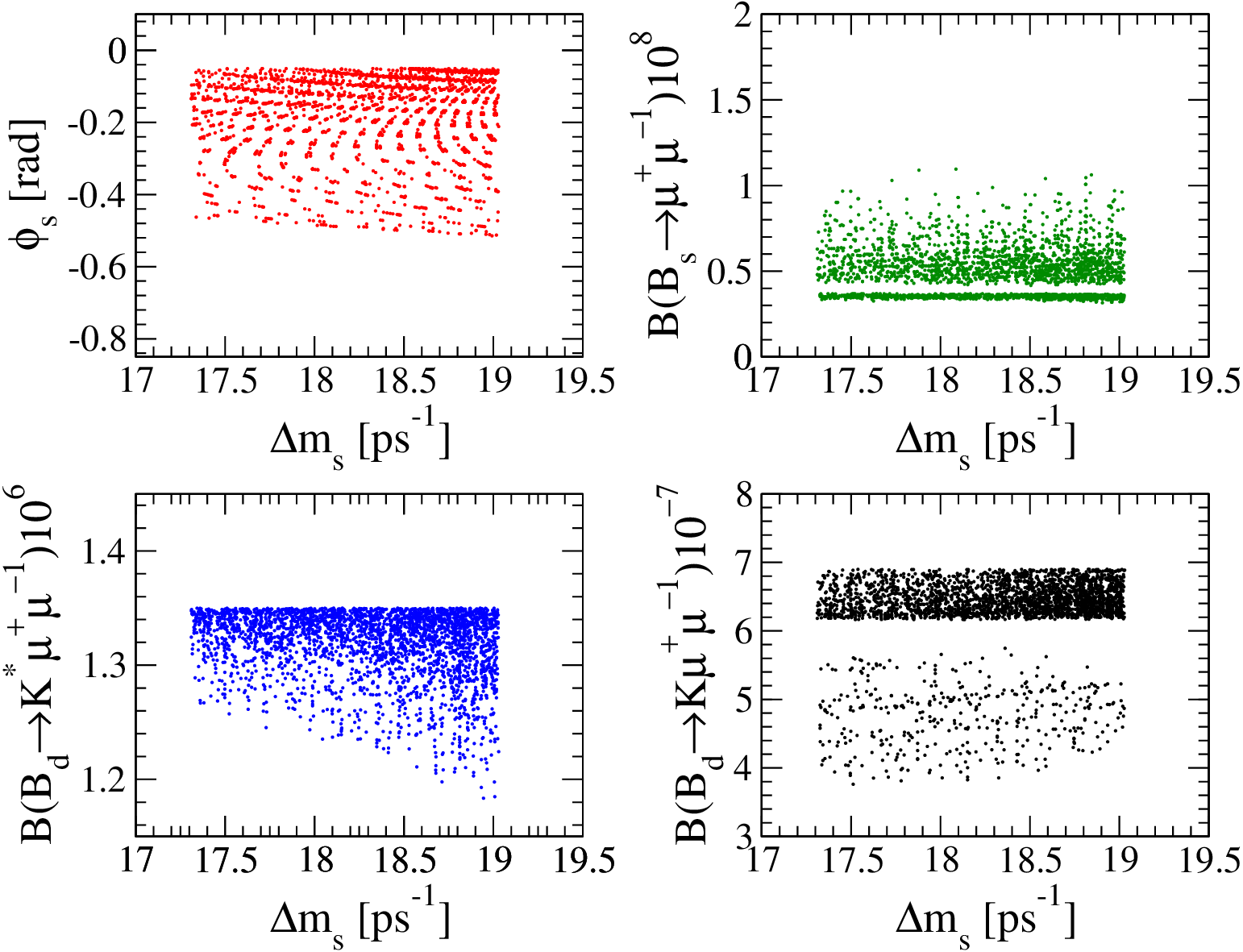}
\caption{Correlations between ($\phi^{J/\Psi \phi}_s$,
 ${\cal B}(\bar B_s\to \mu^{+}\mu^{-})$,
${\cal B}(\bar B_d\to K^{*0}\mu^{+} \mu^{-})$, ${\cal B}(\bar B_d\to
K^{0}\mu^{+} \mu^{-})$) and  $\Delta m_s$ by the influence of
nonuniversal $Z'$.}
 \label{fig:phis}
\end{figure}

\begin{figure}[bpth]
\includegraphics*[width=4 in]{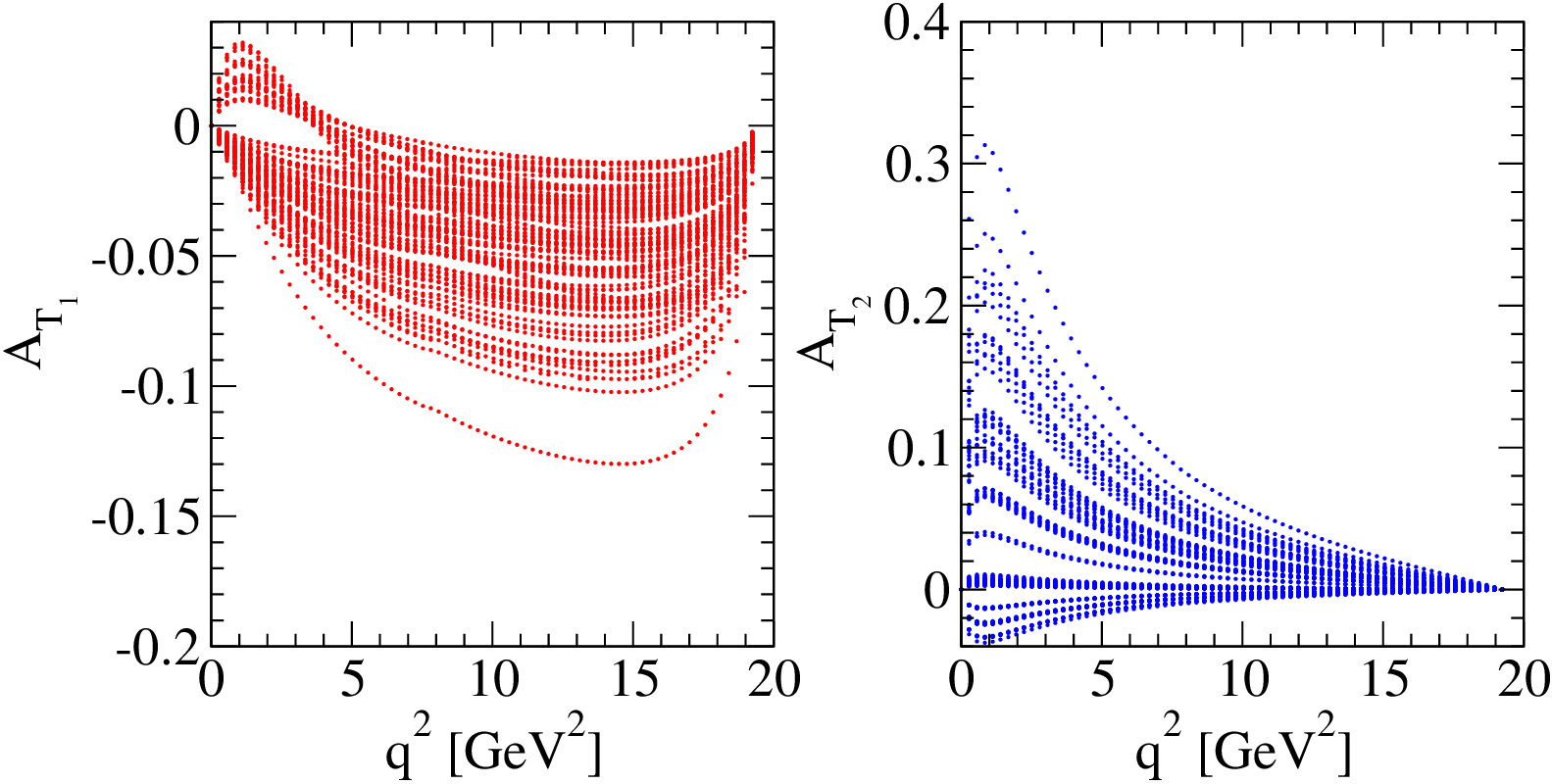}
\caption{T-odd observables ${\cal A}_{T_{1,2}}$ in $\bar B_d\to
K^{*0}\mu^{+} \mu^{-}$ as a function of $q^2$ in nonuniversal $Z'$
model.}
 \label{fig:tv}
\end{figure}

\begin{figure}[tbhp]
\includegraphics*[width=4 in]{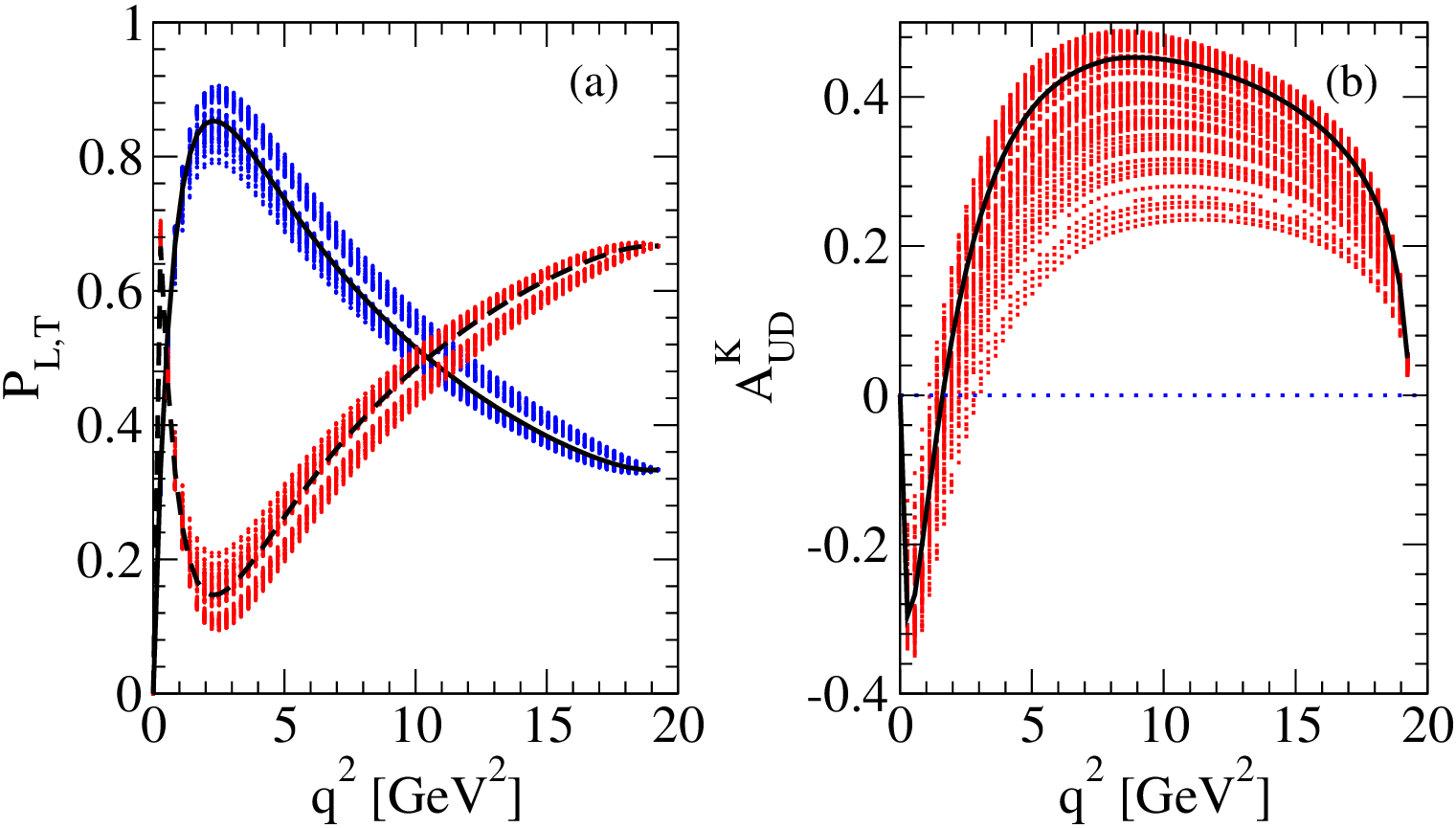}
\caption{ (a) Longitudinal [Transverse] polarization ${\cal
P}_{L[T]}$ of $K^*$ and (b) up-down asymmetry of K in $\bar B_d\to
K^{*0}(\to K \pi) \mu^+ \mu^-$, where solid and dashed lines stand
for the SM predictions. }
 \label{fig:ud}
\end{figure}

We now present the numerical results. In order to realize the strong
correlation among $\phi^{J/\Psi \phi}_{s}$, $\Delta m_s$, ${\cal
B}(\bar B_s\to \mu^{+} \mu^{-})$ and ${\cal B}(\bar B_d\to
K^{[*]0}\ell^{+} \ell^{-})$ influenced by the same parameters, we
show the observables versus $\Delta m_s$  in Fig.~\ref{fig:phis}.
Here, for the $\bar B\to K^{(*)}$ transition form factors, we employ
the results calculated by light-cone sum rules (LCSRs) \cite{LCSR}.
By the figure, we see clearly that besides the BR of $\bar B_s\to
\mu^{+} \mu^{-}$ can be enhanced to be close to $O(10^{-8})$,  the
large $\phi^{J/\Psi \phi}_s$ generated by $Z'$-mediated effects can
be consistent with the current measurements of D{\O} and CDF.
With the values of parameters which are constrained by the current data
of $\Delta m_s$, ${\cal B}(\bar B_s\to \mu^{+} \mu^{-})$ and ${\cal
B}(\bar B_d\to K^{[*]0}\ell^{+} \ell^{-})$, the predicted T-odd
observables of ${\cal A}_{T_{1,2}}$ are displayed in
Fig.~\ref{fig:tv}. It is clear that we can obtain ${\cal A}_{T_1}$
of $10\%$ and ${\cal A}_{T_2}$ of $20\%$ at large and small $q^2$,
respectively.
Moreover, with the same set of values of parameters, the predicted
longitudinal [transverse] polarization and the UD asymmetry defined
in Eq.~(\ref{eq:pl}) [Eq.~(\ref{eq:pt})] and Eq.~(\ref{eq:UDA}) are
given in Fig.~\ref{fig:ud}(a) and (b), respectively, where the solid
and dashed lines denote the SM predictions. We find that the
deviation of 10\% from the SM value in ${\cal P}_{L[T]}$ can be
achieved in the $Z'$ models. Additionally, the crossing of ${\cal
P}_L$ and ${\cal P}_{T}$ at large $q^2$ is also sensitive to the new
physics effects. Besides a large deviation from the SM result in the
spectrum of ${\cal A}^{K}_{UD}$ could be accomplished, the zero
point of ${\cal A}^{K}_{UD}$ is also sensitive to the $Z'$-mediated
effects.

In summary, we have studied the $Z'$-mediated effects on $\Delta
m_s$, $\phi^{J/\Psi \phi}_{s}$, $\bar B_s\to \mu^+ \mu^-$ and $\bar
B\to V \ell^+ \ell^-$ in the family dependent $U(1)'$ model. With
the constraints of the current data, we have found that in the
nonuniversal $Z'$ model, the time-dependent CP violating phase of
the $B_s$ oscillation could be $\phi^{J/\Psi\phi}_s=-(0.26\pm 0.22)$
and the T-odd observables ${\cal A}_{T_1}$ and ${\cal A}_{T_2}$
could be $10\%$ and $20\%$ at the large and small $q^2$ regions,
respectively. In addition, we have also shown that the longitudinal
and transverse polarizations of $K^*$ and the UD asymmetry of $K$ in
the decay chain $\bar B\to K^*(\to K\pi) \ell^+ \ell^-$ could be
significantly influenced by the $Z'$ effects.

\section*{Acknowledgments}
 This work is supported in part by
the National Science Council of R.O.C. under Grant No:
NSC-97-2112-M-006-001-MY3.


\end{document}